%% file: ionization_screening_time.tex
\documentclass[12pt]{iopart}
\usepackage{graphicx}
\usepackage{epstopdf}
\usepackage{array}
\usepackage{enumerate}
\usepackage{url}
\usepackage{amssymb}

\begin{document}

\title[Electrical screening in discharges] {A time scale for electrical
  screening in pulsed gas discharges} \author{Jannis Teunissen$^1$, Anbang
  Sun$^{1}$, Ute Ebert$^{1,2}$}

\address{$^1$Centrum Wiskunde \& Informatica (CWI), P.O.
  Box 94079, 1090 GB Amsterdam, The Netherlands,}
\address{$^2$Dept.\
  Physics, Eindhoven Univ.\ Techn., The Netherlands,}
\ead{jannis@teunissen.net}

\begin{abstract}
  The Maxwell time is a typical time scale for the screening of an
  electric field in a medium with a given conductivity.
  We introduce a generalization of the Maxwell time that is valid for
  gas discharges: the \emph{ionization screening time}, that takes the
  growth of the conductivity due to impact ionization into account.
  We present an analytic estimate for this time scale, assuming a
  planar geometry, and evaluate its accuracy by comparing with
  numerical simulations in 1D and 3D.
  We investigate the minimum plasma density required to prevent the
  growth of streamers with local field enhancement, and we discuss the
  effects of photoionization and electron detachment on ionization
  screening.
  Our results can help to understand the development of pulsed
  discharges, for example nanosecond pulsed discharges at atmospheric
  pressure or halo discharges in the lower ionosphere.
\end{abstract}

\maketitle

\section{Introduction}
\label{sec:introduction}

When a weakly ionized plasma is exposed to an external electric field,
charges will move to screen the plasma interior from the field.
A typical time scale for this process is the Maxwell time, also known
as the dielectric relaxation time \cite{Barnes_1987}, that depends on
the mobility and density of charge carriers in the plasma.
In this paper, we present a generalization of the Maxwell time that is
also valid for electric fields above breakdown, by taking into account
charge multiplication.
We call this generalization the \emph{ionization screening time}.

Our motivation for investigating electric screening in discharges came
from two other articles \cite{Sun_2013, Sun_in_prep_2014}, in which we
simulated the breakdown of ambient air.
We included background ionization in the form of negative ions, from
which electron avalanches could grow after electron detachment.
These avalanches together started screening the electric field, but we
could not simulate up to the end of this process.
Therefore, we briefly introduced the concept of an \emph{ionization
  screening time} in \cite{Sun_2013}.
This name was inspired by a similar phenomenon: after a lightning
stroke, ionization screening waves can form in the lower ionosphere,
also known as halos \cite{Luque_2009}.

In this paper, we investigate the ionization screening time in more
detail.
The paper is organized in the following way.
In section \ref{sec:screening-time} the Maxwell time is discussed and
the ionization screening time is introduced.
Our analytic estimate for the ionization screening time is compared
with simulation results in section \ref{sec:comparison-simulations}.
These simulations are performed in 1D and 3D, using a fluid and a
particle model.
For low levels of initial ionization, discharges become inhomogeneous
and local field enhancement becomes important, which is investigated
in section \ref{sec:homogeneity}.
Finally, we discuss the effect of electron detachment and
photoionization on the screening process in section \ref{sec:other},
which is especially relevant for air.

\section{The ionization screening time}
\label{sec:screening-time}

Below, we first discuss the Maxwell time, also known as the
dielectric relaxation time \cite{Barnes_1987}.
Then we introduce the ionization screening time, for which we give an
analytic estimate.

\subsection{The Maxwell Time}
\label{sec:maxwell-time}

Although the Maxwell time is valid for any medium with a constant
conductivity, we focus here on the case of a plasma.
Suppose we have a neutral plasma with an electron density $n_e$ on which an
electric field $\vec{E}$ is applied.
The field accelerates the electrons in the plasma, while collisions slow them
down again.
This gives rise to an electrical current
\begin{equation}
  \label{eq:current_assumption}
  \vec{J}_e = e n_e \mu_e \vec{E},
\end{equation}
where $\mu_e$ denotes the electron mobility and $e$ the elementary charge.
(We ignore the much smaller contribution of the ions.)
This current reduces the electric field inside the plasma.
By taking the divergence of Amp\`{e}re's law, we can relate the current to the
time derivative of the electric field
\begin{equation}
  \label{eq:ampere_law}
  \nabla \cdot \left(\vec{J}_e + \varepsilon_0 \partial_t \vec{E}\right) = 0,
\end{equation}
where $\varepsilon_0$ is the dielectric permittivity.
This equation can be interpreted more easily if we assume the system
is planar, i.e., effectively one-dimensional, so that we get a scalar
equation.
If a constant external field $E_0$ (i.e., $\partial_t E_0 = 0$) is
applied from some location outside the plasma, integration of
(\ref{eq:ampere_law}) gives
\begin{equation}
  \partial_t E = - J_e/\varepsilon_0
  = - (e n_e \mu_e / \varepsilon_0) \, E.
\end{equation}
A typical time scale for electric screening is given by
$-E/\partial_t E$, which is called the Maxwell time:
\begin{equation}
  \label{eq:maxwell_time}
  \tau_\mathrm{Maxwell} = \varepsilon_0 / (e n_e \mu_e).
\end{equation}
For a different derivation see \cite{surzhikov2012computational}.
Note that there is no dependence on the density profile at the plasma
boundary.

\subsection{The ionization screening time}
\label{sec:ionization_screening-time}

When the field $E_0$ applied to a plasma is above the breakdown
threshold, the Maxwell time is no longer valid, because the electron
density $n_e$ grows in time.

We present a generalization of the Maxwell time, which we call the \emph{ionization
screening time} or $\tau_\mathrm{is}$.
It estimates how long it takes for the electric field inside a discharge to drop
below the breakdown threshold.
Below we present a derivation, the result of which is
\begin{equation}
  \label{eq:ionization_screening-time}
  \tau_\mathrm{is} =
  \ln\left(1+\frac{\alpha_\mathrm{eff} \varepsilon_0 E_0}{e n_0}\right) /
  (\alpha_\mathrm{eff} \mu_e E_0),
\end{equation}
where $\alpha_\mathrm{eff}$ is the effective ionization coefficient.
Note that in the limit $\alpha_\mathrm{eff} \to 0$, equation
(\ref{eq:ionization_screening-time}) reduces to the Maxwell time
(\ref{eq:maxwell_time}).

\subsection{Analytic estimate}
\label{sec:estimate-screening-time}

To derive an analytic estimate for the ionization screening time, we study a
simplified system. The assumptions are listed below:
\begin{itemize}
  \item The system is planar (effectively one-dimensional);
  there is spatial variation in the $x$-direction only.
  \item Initially, the electron and ion density is $n_0$ between $x_0$ and
  $x_1$, and zero elsewhere.
  The width $x_1 - x_0$ is taken larger than the distance the electrons will
  drift up to the ionization screening time.
  \item Electrons keep the same drift velocity $v_d = \mu_e E_0$ and effective
  ionization coefficient $\alpha_\mathrm{eff}$ as in the initial background
  field $E_0$.
  \item There is no diffusion.
\end{itemize}

\begin{figure}
  \begin{center}
    \includegraphics[width=10cm]{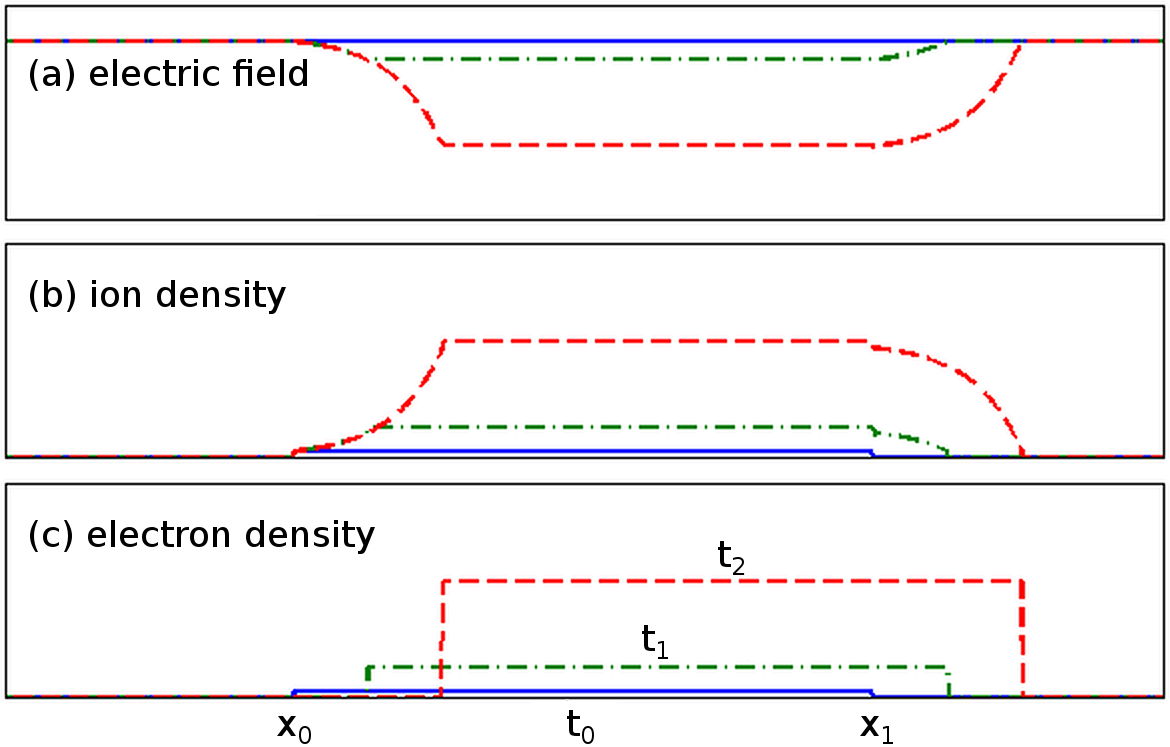}
    \caption{Schematic view of (a) electric field, (b) ion density and
      (c) electron density at three times $t_0 < t_1 < t_2$.
      The electric field decreases in the ionized region due to the charge
      separation at the left and right boundary.}
    \label{fig:screening-time_domain}
  \end{center}
\end{figure}

The evolution of this system will resemble the one depicted in
figure~\ref{fig:screening-time_domain}.
The electrons, which are initially present between $x_0$ and $x_1$,
drift to the right with velocity $v_d$.
Their number density grows in time as $e^{\alpha_\mathrm{eff} v_d t}$.
At time $t$ there are no electrons below $x_0 + v_d t$, while they
have created an ion density $n_0 e^{\alpha_\mathrm{eff} (x-x_0)}$
between $x_0$ and $x_0 + v_d t$.
Therefore, the integrated net charge in this region is
$(e^{\alpha_\mathrm{eff} v_d t}-1) e n_0 /\alpha_\mathrm{eff}$.
Equating this to the charge $\varepsilon_0 E_0$ needed to screen an
electric field $E_0$, and solving for $t$ gives the following
expression for the ionization screening time
\begin{equation}
  \tau_\mathrm{is} =
  \ln\left(1+\frac{\alpha_\mathrm{eff} \varepsilon_0 E_0}{e n_0}\right) /
  (\alpha_\mathrm{eff} v_d),
  \label{eq:screening-time}
\end{equation}
where $v_d$ can be replaced by $\mu_e E_0$.

In deriving equation (\ref{eq:screening-time}) we have assumed that
$\alpha_\mathrm{eff}$ and $v_d$ keep their values for the initial field $E_0$.
This approximation becomes more accurate if the initial electron density $n_0$
is small compared to the density at the time of screening.
Then the electric field stays close to $E_0$ during most of the screening
process, because the charge density is not yet large enough to affect it.
Note that by using these initial coefficients we will underestimate the
ionization screening time.
This is somewhat compensated for by computing the time to shield the electric
field to zero, instead of to a value below breakdown.

\section{Comparison with simulations}
\label{sec:comparison-simulations}

We will now compare the predictions of
equation~(\ref{eq:screening-time}) with numerical simulations.
In these simulations, we determine how long it takes for the electric
field inside a discharge to drop below the breakdown threshold.
We perform these comparisons in nitrogen at 1 bar and 293 Kelvin, for
which we have used a breakdown field of $3 \, \textrm{MV/m}$.
(Since there are no electron loss mechanisms in pure nitrogen, the
breakdown field is not well-defined.)
Below, we describe the simulation models.

\subsection{Simulation Models}
\label{sec:sim-models}

We use two types of simulation models here: a plasma fluid model (1D)
and a particle model (1D and 3D).
It will turn out that in 1D, the fluid model gives almost the same
results as the particle model.
We nevertheless include both, to provide a link between the 3D
particle simulations presented in section \ref{sec:comparison_3d} and
the plasma fluid description used for equation
(\ref{eq:screening-time}).

In all cases, a spatial resolution of $8 \, \mu\textrm{m}$ and
a time step of $1 \, \textrm{ps}$ was used.
In 1D, the computational domain was $16 \, \textrm{mm}$ long.
In 3D, the computational domain measured 8 mm along the $x$-direction,
with an area of $1\times 1$ mm$^2$ in the transverse direction.
To get the planar structure of the 1D simulations in 3D, we have used
periodic boundary conditions in the transverse direction.

\subsubsection{1D fluid model}
\label{sec:model_1d_fluid}

The plasma fluid model that we use is of the drift-diffusion-reaction type \cite{Montijn_2006}.
It contains the following equations:
\begin{eqnarray}
  \partial_t n_e &=
  \nabla \cdot (\mu_e \vec{E} n_e + D_e \nabla n_e)
  + \alpha_\mathrm{eff} \mu_e |\vec{E}| n_e,\\
  \partial_t n^{+} &= \alpha_\mathrm{eff}  \mu_e |\vec{E}| n_e,\\
  \nabla\cdot\vec{E} &= e (n^{+}-n_e)/\varepsilon_0,
  \label{eq:simple_fluid_equations}
\end{eqnarray}
where $D_e$ is electron diffusion coefficient and $n^{+}$ is the
density of positive ions.
In the simulations, the coefficients $\mu_e$, $D_e$ and
$\alpha_\mathrm{eff}$ depend on the local electric field, which is
recomputed at every time step.
These coefficients are computed from the particle cross sections
\cite{siglo_database} by measuring the properties of simulated
particle swarms, see \cite{Raspopovic_1999}.
The same coefficients are used for equation (\ref{eq:screening-time}).

The fluid equations are solved with a third-order upwind scheme, as in
\cite{Montijn_2006}.
Time stepping was done with the classic fourth order Runge-Kutta scheme.

\subsubsection{3D particle model}
\label{sec:model_3d_particle}

The 3D model is of the PIC-MCC type, with electrons as particles and
ions as densities.
The electrons randomly collide with a background of neutral molecules.
We use cross sections from the Siglo database \cite{siglo_database},
Fishpack \cite{fishpack90} to compute the electric potential and
adaptive particle management for the super-particles
\cite{Teunissen_2014}.
This model is described in some detail in \cite{Sun_2013,
  Sun_in_prep_2014}.

\subsubsection{1D particle model}
\label{sec:model_1d_particle}

The 1D particle model was constructed from the 3D particle model described above.
The 3D model is converted to 1D by projecting the particles onto one spatial
dimension for the calculation of the electric field.
The particles then have just one coordinate for their position, but their
velocities still have three components.

\subsection{Comparison with 1D simulations}
\label{sec:comparison_results}

We now compare our analytic approximation to the two numerical
simulation models.
In figure \ref{fig:ist_comp_1d}, we show the screening time for fields
between 5 and 10 MV/m.
Two initial conditions are used: an electron and ion density of
$10^{13}$ or $10^{11} \, \mathrm{m}^{-3}$ was present between 12 and
14 mm.
Equation (\ref{eq:screening-time}) predicts shorter screening times
than we see in the simulations, but the agreement is nevertheless
quite good.
Note that the particle and fluid model give almost identical results.

\begin{figure}
  \begin{minipage}{0.45\textwidth}
    \centering
    \footnotesize
    \input{fig-partial-screening-field_pdf.tex}
  \end{minipage}
  \hspace{0.05\textwidth}
  \begin{minipage}{0.45\textwidth}
    \centering
    \footnotesize
    \input{fig-partial-screening-dens_pdf.tex}
  \end{minipage}
  \caption{Partial screening of the electric field in the 1D fluid simulations,
    for a background field of 6 MV/m and an initial plasma density of $10^{13}$ m$^{-3}$.
    The electric field (left) and the electron density (right) are shown at
    various times. The exponential growth of the electron density slows down
    because the electric field gets screened.
    }
  \label{fig:partial_screening}
\end{figure}
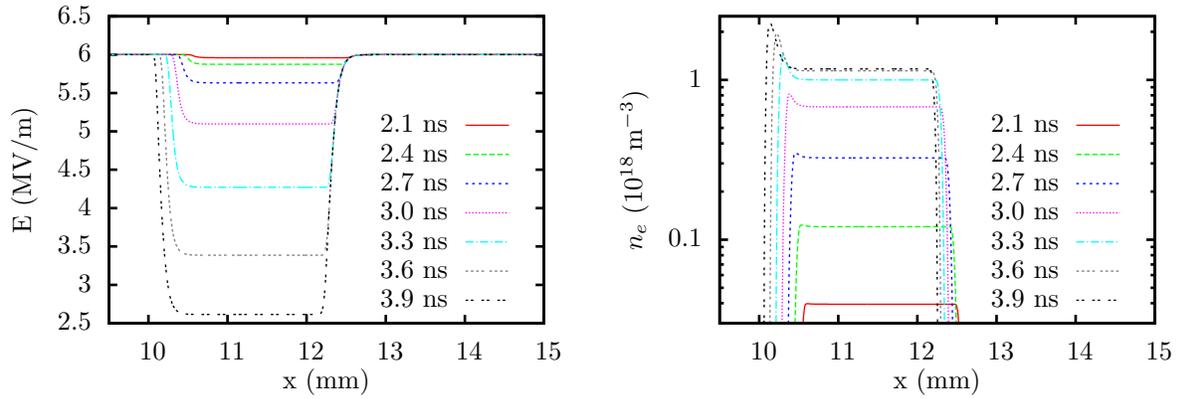

As discussed in section \ref{sec:estimate-screening-time}, the partial
screening of the electric field was not included in deriving equation
(\ref{eq:screening-time}).
An example of this partial screening is shown in figure
\ref{fig:partial_screening}, where the electric field and the electron
density are shown at various times, using the 1D fluid model in a
background field of 6 MV/m.
Close to the screening time, the exponential growth of the electron
density slows down, because the field is partially screened.

\begin{figure}
  \centering
  \footnotesize
  \input{fig-ist-1d_pdf.tex}
  \caption{The ionization screening time versus the applied electric field, for two initial plasma densities ($10^{11}$ and $10^{13}$).
    Results are shown for a 1D fluid model, a 1D particle model and
    equation (\ref{eq:screening-time}), for N$_2$ at 1 bar.
    }
  \label{fig:ist_comp_1d}
\end{figure}
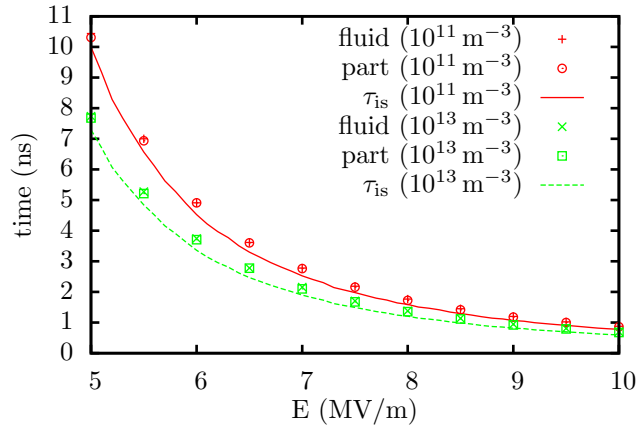

\subsection{Comparison with 3D simulations}
\label{sec:comparison_3d}

To investigate how inhomogeneities affect the ionization screening
time, we have performed 3D particle simulations in a field of 6 MV/m.

We will show results using two initial plasma densities: $10^{13}$ and
$10^{11}$ m$^{-3}$.
In both cases, the plasma is initially present between 4 and 6 mm.
Because the electric field is now a varying 3D vector field, we cannot
directly compare it to the 1D results.
Therefore, we show the electric field and the electron density
averaged over transverse planes.
This leaves only the longitudinal component of the field nonzero, due
to the periodic boundary conditions.

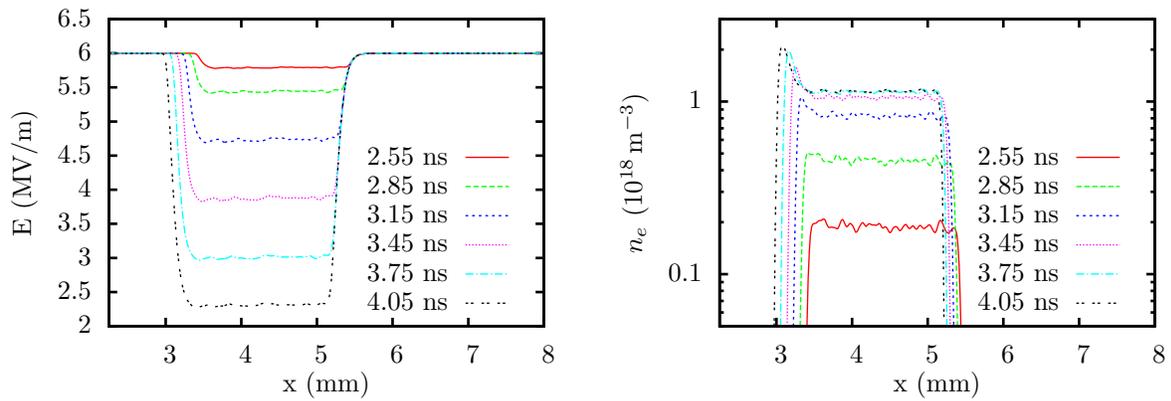
\begin{figure}
  \begin{minipage}{0.45\textwidth}
    \centering
    \footnotesize
    \input{fig-3d-screening-1e13-field_pdf.tex}
  \end{minipage}
  \hspace{0.05\textwidth}
  \begin{minipage}{0.45\textwidth}
    \centering
    \footnotesize
    \input{fig-3d-screening-1e13-dens_pdf.tex}
  \end{minipage}
  \caption{Electric field and electron density in the 3D particle simulations,
    for a background field of 6 MV/m and an initial density of $n_0 =
    10^{13}\,\mathrm{m}^{-3}$. The values are averaged over planes perpendicular
    to the $x$-direction.
  }
  \label{fig:screening-3d-1e13}
\end{figure}
\begin{figure}
  \centering
  \includegraphics[width=10cm]{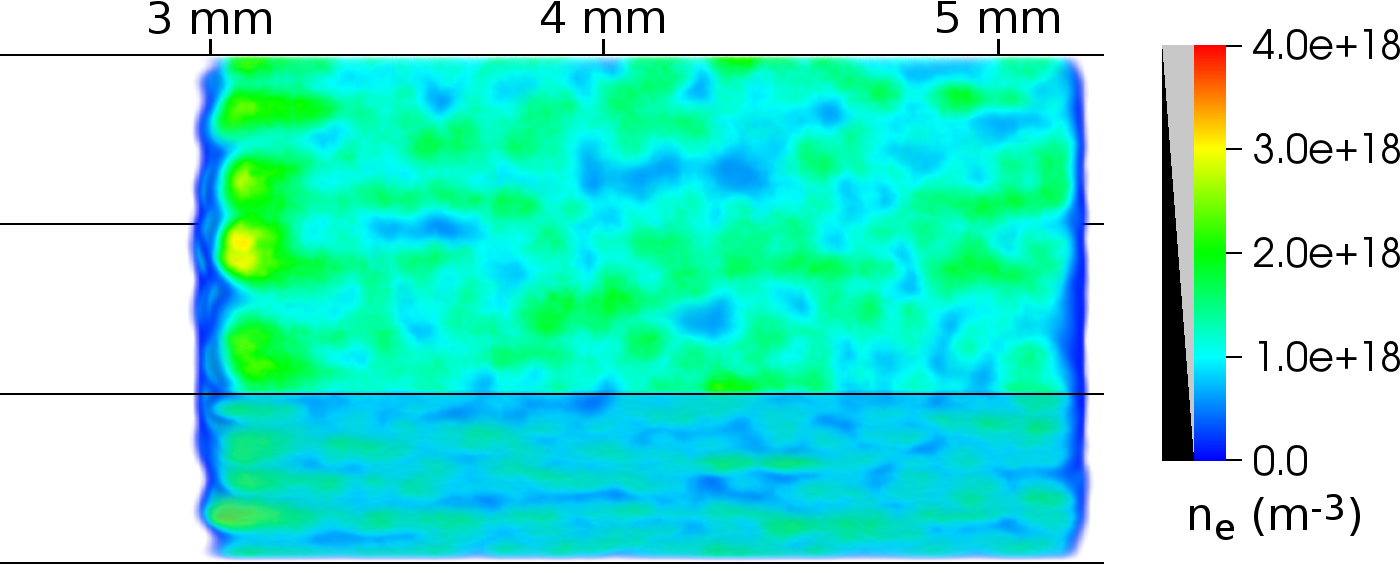}
  \caption{The electron density in the 3D particle model at $4.05 \,
    \textrm{ns}$, for an initial density of $n_0 = 10^{13}\,\mathrm{m}^{-3}$.
    (This figure is made using volume rendering; transparency is indicated in the legend.)}
  \label{fig:elec_dens_3d_1e13}
\end{figure}

We first present the results for an initial density of $n_0 = 10^{13}
\, \textrm{m}^{-3}$ between 4 and 6 mm.
In figure \ref{fig:screening-3d-1e13} we present averaged electric
field and electron density profiles at various times.
In figure \ref{fig:elec_dens_3d_1e13}, a 3D view of the electron
density at $4.05 \, \textrm{ns}$ is shown.
The screening time is about $3.75 \, \textrm{ns}$, as in the 1D case
of figure \ref{fig:partial_screening}.
Some noise can be seen in the electric field and density profiles,
because the initial density corresponds to $10^4$ electrons per
mm$^3$.

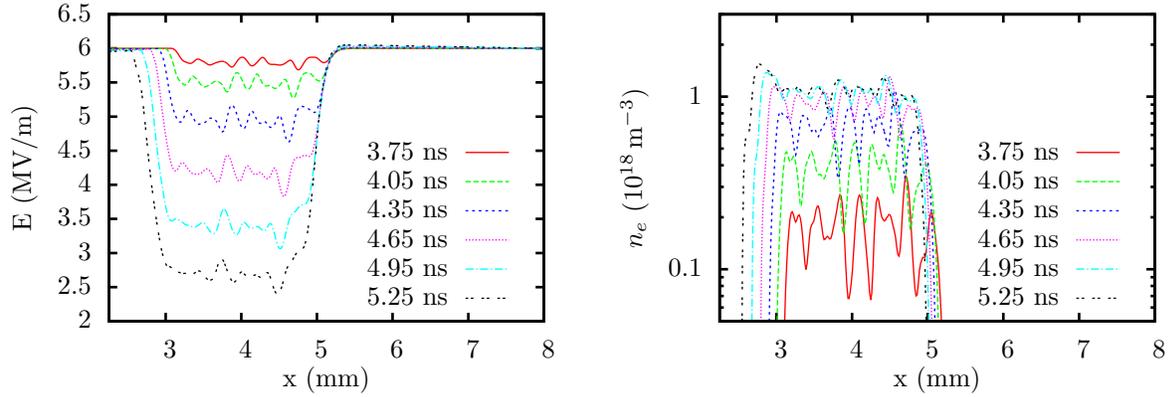
\begin{figure}
  \begin{minipage}{0.45\textwidth}
    \centering
    \footnotesize
    \input{fig-3d-screening-1e11-field_pdf.tex}
  \end{minipage}
  \hspace{0.05\textwidth}
  \begin{minipage}{0.45\textwidth}
    \centering
    \footnotesize
    \input{fig-3d-screening-1e11-dens_pdf.tex}
  \end{minipage}
  \caption{Electric field and electron density in the 3D particle simulations, as
    in figure \ref{fig:screening-3d-1e13}, but now for a lower initial density
    of $n_0 = 10^{11}\,\mathrm{m}^{-3}$.
  }
  \label{fig:screening-3d-1e11}
\end{figure}
\begin{figure}
  \centering
  \includegraphics[width=10cm]{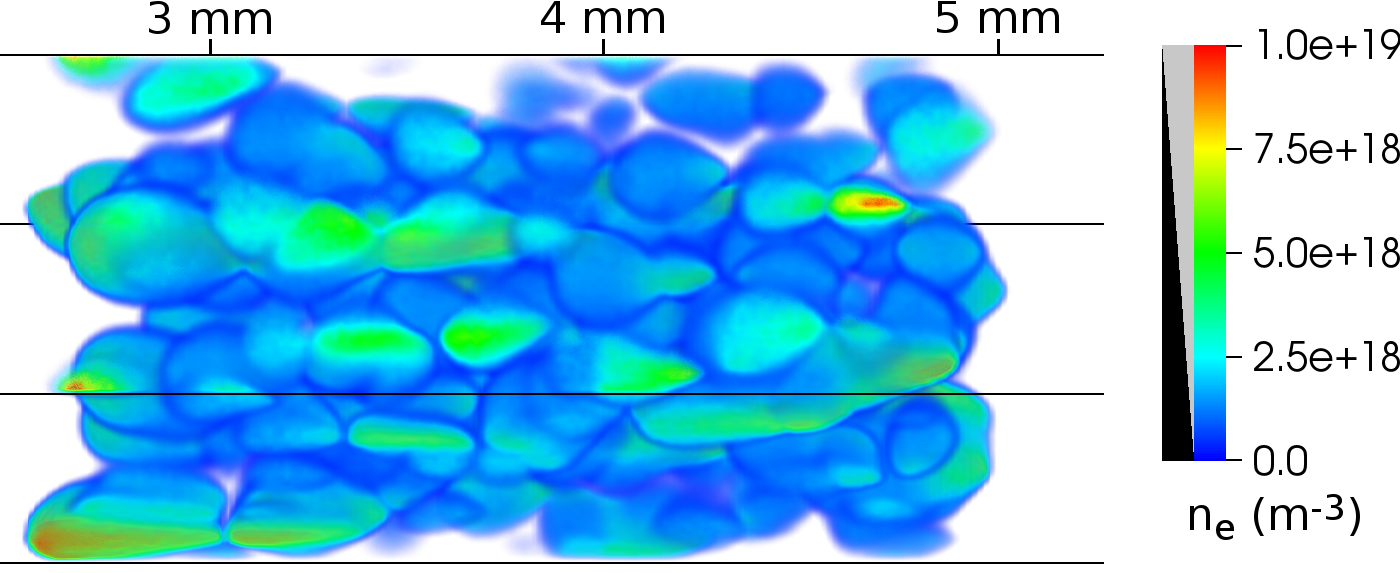}
  \caption{The electron density in the 3D particle model at $5.25 \,
    \textrm{ns}$, for an initial density of $n_0 = 10^{11}\,\mathrm{m}^{-3}$.}
  \label{fig:elec_dens_3d_1e11}
\end{figure}

With an initial density of $n_0 = 10^{11} \, \textrm{m}^{-3}$, the
results look quite different, see figures \ref{fig:screening-3d-1e11}
and \ref{fig:elec_dens_3d_1e11}.
There is now significant noise in the electric field and especially in
the electron density profiles.
These larger fluctuations emerge because the initial density
corresponds to only $10^2$ electrons per mm$^3$.
The screening time is about $5.1 \, \textrm{ns}$, which is still in
agreement with the 1D results of figure \ref{fig:ist_comp_1d}.

Compared to the 1D results, we observe almost the same screening times
in 3D, but with lower initial densities fluctuations become larger.
If we would further reduce the initial electron density, we would
eventually get a few separated electron avalanches that develop into
streamers.

\section{The homogeneity of discharges}
\label{sec:homogeneity}

In the previous section we have seen that discharges can develop quite
irregularly if the initial electron density is low.
The irregularities cause field enhancement, that could invalidate our estimate
for the ionization screening time.
To estimate when this happens, we first discuss how long it takes for space
charge effects to develop.

\subsection{The streamer formation time}
If an electron avalanche starts from a single electron, how long does
it take for space charge effects to become significant?
In other words, how long does it take for a streamer to form?
The answer depends on the proccesses that can affect the space charge
fields: ionization, drift and diffusion.
The coefficients of these processes can be described in terms of the
electric field $E$ and the gas number density $N$, so that in general
the `streamer formation time' is a function of $E$ and $N$.
According to~\cite{Montijn_2006_diff, Ebert_2010}, the number of
electrons required for a streamer to form scales as $g(E) \cdot N_0/N$,
where $g(E)$ is some function of the electric field and $N_0$ is the
density of air at standard temperature and pressure.
Then the time scale for streamer formation can be expressed as
\begin{equation*}
  \tau_\mathrm{streamer}=\ln \left[ g(E) \cdot N_0/N \right]
  /(\alpha_\mathrm{eff} v_d).
\end{equation*}
For $N=N_0$, a commonly used empirical approximation is to take $g(E)
\approx 10^8$, so that $\ln[g(E)] \approx 18$.
This criterion is know as the Raether-Meek criterion, for which the
streamer formation time is given by
\begin{equation}
  \tau_{\rm RM}\approx18 / (\alpha_\mathrm{eff} v_d).\label{eq:streamer_time}
\end{equation}

\subsection{Required pre-ionization for homogeneity}
\label{sec:req-homogeneity}

From the previous section we have an estimate for the time it takes to
develop space charge effects.
Given this time, we can estimate how high the initial electron density
$n_0$ needs to be to prevent streamer formation.
Several authors have made such estimates in the past, see for example
\cite{Palmer_1974,Levatter_1980,Herziger_1981}.
Much of this research was aimed at generating homogeneous discharges
for CO$_2$ lasers.
Below, we derive an alternative criterion for homogeneity that is
based on arguments from
\cite{Palmer_1974,Levatter_1980,Herziger_1981}, but perhaps simpler.

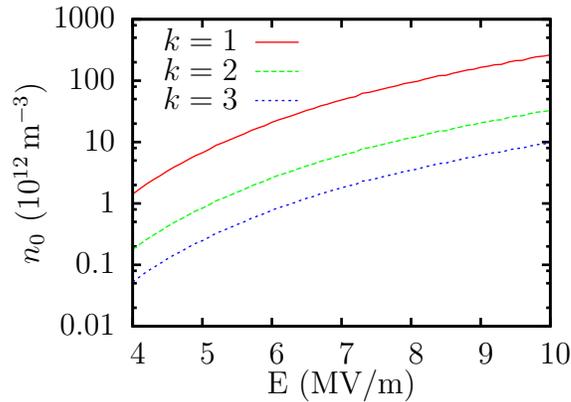
\begin{figure}
  \centering
  \input{fig-n0-min_pdf.tex}
  \caption{The required initial electron density for homogeneous breakdown
    according to equation (\ref{eq:required_dens}), for three values
    of $k$. The curves shown are for N$_2$ at 1 bar.
  }
  \label{fig:required_n0}
\end{figure}

As long as space charge effects are negligible, the electron avalanche
will radially expand due to diffusion.
In the radial direction, the electron density at time $t$ will have a
Gaussian distribution with a standard deviation of $\sqrt{2 D_e t}$.
If we let $R_s$ denote the typical radius at the time of streamer
formation, see equation~(\ref{eq:streamer_time}), we get $R_s = 6
\sqrt{D_e/(v_d \alpha)}$.
If streamer formation is to be prevented, the avalanches need to be
sufficiently close to each other.
This means that their initial separation should be on the order of
$R_s$.
Suppose it is $k \cdot R_s$, where $k$ is about one, then the initial
electron density $n_0$ should be at least
\begin{equation}
  n_0 \approx 1/(R_s)^3 = \frac{1}{216 k^3}\,\left(\frac{v_d \alpha} {D_e}\right)^{3/2}.
  \label{eq:required_dens}
\end{equation}
In figure \ref{fig:required_n0}, equation (\ref{eq:required_dens}) is shown
against the electric field for N$_2$ at 1 bar, for three values of $k$.
We can see that the result is quite sensitive to $k$.
Using $1 \leq k \leq 3$, the required initial density lies between
$10^{12}$ and $10^{13} \, \textrm{m}^{-3}$ for a field of $6 \, \textrm{MV/m}$,
in agreement with the results from section \ref{sec:comparison_3d}.

\section{The effect of detachment and photoionization}
\label{sec:other}

Besides impact ionization, there can be other ways to generate free electrons in
a gas, which may affect our estimate for the screening time.
This is especially true for air, in which electron detachment and
photoionization can occur.
The effect of these processes is discussed below.

\subsection{Electron detachment}
\label{sec:screening-time_delay}

In electronegative gases there might initially be negative ions
instead of free electrons.
Ionization screening by electrons can still take place in such a gas
if electrons are able to detach from the negative ions.
If a typical time scale for detachment is $\tau_d$, then the screening
process will be delayed by approximately
\begin{equation}
  \label{eq:tau-delay}
  \tau_\mathrm{delay} = \ln(1+\tau_d \alpha_\mathrm{eff} v_d) / (\alpha_\mathrm{eff} v_d),
\end{equation}
so that the total screening time is given by the sum of equations
(\ref{eq:screening-time}) and (\ref{eq:tau-delay})
\begin{equation}
  \label{eq:screening-time-det}
  \tau_\mathrm{is} =
  \left[
    \ln\left(1+\frac{\alpha_\mathrm{eff} \varepsilon_0 E_0}{e n^{-}}\right)
    + \ln(1+\tau_d \alpha_\mathrm{eff} v_d)
  \right]
  / (\alpha_\mathrm{eff} v_d),
\end{equation}
where $n^{-}$ denotes the negative ion density.
Equation (\ref{eq:tau-delay}) is the solution to
$n_e(\tau_\mathrm{delay}) = n^{-}$ given the following equation
\begin{equation*}
  \partial_t n_e(t) = n^{-} / \tau_d + n_e(t) \alpha_\mathrm{eff} v_d,
\end{equation*}
with $n_e(0) = 0$.
This last equation describes the growth of the electron density in time, but it
does not take the depletion of negative ions by detachment into account ($n^{-}$
should change in time).
The underlying assumption is that ionization quickly dominates over detachment.
Furthermore, the coefficients $\tau_d$, $\alpha_\mathrm{eff}$ and $v_d$ are
assumed to be constant, since we do not expect the electric field to change
during the detachment phase.

Summarizing, if there are negative ions from which electrons first have to
detach, then there will be a delay in the ionization screening process. The
ionization screening time can then be approximated by equation (\ref{eq:screening-time-det}).

\subsection{Photoionization}
\label{sec:photoionization}

Photoionization can occur if excited molecules (or atoms) emit photons
energetic enough to ionize other molecules (or atoms).
With a few assumptions, we can estimate how photoionization will
affect the screening time.
Suppose that on average $\eta$ photoionization events take place per
electron-impact ionization.
Suppose further that these photoionizations take place at a distance
that is larger than $n_0^{-1/3}$, where $n_0$ is the initial density
of electrons.
If there is no delay in emitting the ionizing photons, and if space
charge effects can be neglected, then the electron density will grow
as
\begin{equation}
  \label{eq:photo-effect}
  n_e(t)
  = n_0 e^{(1+\eta) \alpha_\mathrm{eff} v_d t}.
\end{equation}
So, photoionization effectively increases $\alpha_\mathrm{eff}$ with a
factor $1+\eta$.
For air at atmospheric pressure $\eta$ is less than 1\%, and at low
pressures $\eta \lesssim 0.1$ \cite{zheleznyak1982}.
Therefore photoionization does not change the ionization screening
time (\ref{eq:screening-time}) much.

Another effect of photoionization could be to make a discharge more
homogeneous.
One interpretation of equation (\ref{eq:photo-effect}) is that
photoionization has effectively increased the initial density $n_0$ by
a factor $e^{\eta \alpha_\mathrm{eff} v_d t}$ at time $t$.
From equation~(\ref{eq:streamer_time}), we get that
$\alpha_\mathrm{eff} v_d t \approx 18$ when space charge effects set
in.
For $\eta = 1\%$, the factor $e^{18 \eta}$ is about $1.2$, so that the
effect of photoionization on the homogeneity of a discharge is rather
weak.
For $\eta \approx 0.1$, the factor is about 6, so that photoionization
should be taken into account.

\section{Conclusion}

We have introduced the ionization screening time, a generalization of
the Maxwell time that is also valid for electric fields above
breakdown.
An analytic estimate for this time scale was introduced, which was
compared with numerical simulations in 1D and 3D, finding good
agreement.
We have given an estimate for the required plasma density to prevent the growth
of inhomogeneities, and we have discussed the effects of electron
detachment and photoionization on ionization screening.

These results can help to understand the development of pulsed
discharges, such as nanosecond pulsed discharges at atmospheric
pressure or halo discharges in the lower ionosphere.
First, our estimate can be used to predict whether such a discharge
initially develops homogeneously.
If so, then two stages can be distinguished: Before the ionization
screening time, growth takes place in the complete discharge volume.
After this time, the discharge grows at its boundary, because its
interior is electrically screened.

\ack{JT was supported by STW project 10755.
  ABS acknowledges the support by an NWO Valorization project at CWI and by STW
  projects 10118 and 12119.}

\section*{References}
\bibliographystyle{unsrt}
\bibliography{ionization_screening_time}

\end{document}

%% file: fig-ist-1d_pdf.tex
\begingroup
  \makeatletter
  \providecommand\color[2][]{%
    \GenericError{(gnuplot) \space\space\space\@spaces}{%
      Package color not loaded in conjunction with
      terminal option `colourtext'%
    }{See the gnuplot documentation for explanation.%
    }{Either use 'blacktext' in gnuplot or load the package
      color.sty in LaTeX.}%
    \renewcommand\color[2][]{}%
  }%
  \providecommand\includegraphics[2][]{%
    \GenericError{(gnuplot) \space\space\space\@spaces}{%
      Package graphicx or graphics not loaded%
    }{See the gnuplot documentation for explanation.%
    }{The gnuplot epslatex terminal needs graphicx.sty or graphics.sty.}%
    \renewcommand\includegraphics[2][]{}%
  }%
  \providecommand\rotatebox[2]{#2}%
  \@ifundefined{ifGPcolor}{%
    \newif\ifGPcolor
    \GPcolortrue
  }{}%
  \@ifundefined{ifGPblacktext}{%
    \newif\ifGPblacktext
    \GPblacktexttrue
  }{}%
  \let\gplgaddtomacro\g@addto@macro
  \gdef\gplbacktext{}%
  \gdef\gplfronttext{}%
  \makeatother
  \ifGPblacktext
    \def\colorrgb#1{}%
    \def\colorgray#1{}%
  \else
    \ifGPcolor
      \def\colorrgb#1{\color[rgb]{#1}}%
      \def\colorgray#1{\color[gray]{#1}}%
      \expandafter\def\csname LTw\endcsname{\color{white}}%
      \expandafter\def\csname LTb\endcsname{\color{black}}%
      \expandafter\def\csname LTa\endcsname{\color{black}}%
      \expandafter\def\csname LT0\endcsname{\color[rgb]{1,0,0}}%
      \expandafter\def\csname LT1\endcsname{\color[rgb]{0,1,0}}%
      \expandafter\def\csname LT2\endcsname{\color[rgb]{0,0,1}}%
      \expandafter\def\csname LT3\endcsname{\color[rgb]{1,0,1}}%
      \expandafter\def\csname LT4\endcsname{\color[rgb]{0,1,1}}%
      \expandafter\def\csname LT5\endcsname{\color[rgb]{1,1,0}}%
      \expandafter\def\csname LT6\endcsname{\color[rgb]{0,0,0}}%
      \expandafter\def\csname LT7\endcsname{\color[rgb]{1,0.3,0}}%
      \expandafter\def\csname LT8\endcsname{\color[rgb]{0.5,0.5,0.5}}%
    \else
      \def\colorrgb#1{\color{black}}%
      \def\colorgray#1{\color[gray]{#1}}%
      \expandafter\def\csname LTw\endcsname{\color{white}}%
      \expandafter\def\csname LTb\endcsname{\color{black}}%
      \expandafter\def\csname LTa\endcsname{\color{black}}%
      \expandafter\def\csname LT0\endcsname{\color{black}}%
      \expandafter\def\csname LT1\endcsname{\color{black}}%
      \expandafter\def\csname LT2\endcsname{\color{black}}%
      \expandafter\def\csname LT3\endcsname{\color{black}}%
      \expandafter\def\csname LT4\endcsname{\color{black}}%
      \expandafter\def\csname LT5\endcsname{\color{black}}%
      \expandafter\def\csname LT6\endcsname{\color{black}}%
      \expandafter\def\csname LT7\endcsname{\color{black}}%
      \expandafter\def\csname LT8\endcsname{\color{black}}%
    \fi
  \fi
  \setlength{\unitlength}{0.0500bp}%
  \begin{picture}(5102.00,3400.00)%
    \gplgaddtomacro\gplbacktext{%
      \csname LTb\endcsname%
      \put(594,594){\makebox(0,0)[r]{\strut{} 0}}%
      \put(594,825){\makebox(0,0)[r]{\strut{} 1}}%
      \put(594,1056){\makebox(0,0)[r]{\strut{} 2}}%
      \put(594,1287){\makebox(0,0)[r]{\strut{} 3}}%
      \put(594,1518){\makebox(0,0)[r]{\strut{} 4}}%
      \put(594,1749){\makebox(0,0)[r]{\strut{} 5}}%
      \put(594,1980){\makebox(0,0)[r]{\strut{} 6}}%
      \put(594,2211){\makebox(0,0)[r]{\strut{} 7}}%
      \put(594,2442){\makebox(0,0)[r]{\strut{} 8}}%
      \put(594,2673){\makebox(0,0)[r]{\strut{} 9}}%
      \put(594,2904){\makebox(0,0)[r]{\strut{} 10}}%
      \put(594,3135){\makebox(0,0)[r]{\strut{} 11}}%
      \put(726,374){\makebox(0,0){\strut{} 5}}%
      \put(1522,374){\makebox(0,0){\strut{} 6}}%
      \put(2318,374){\makebox(0,0){\strut{} 7}}%
      \put(3113,374){\makebox(0,0){\strut{} 8}}%
      \put(3909,374){\makebox(0,0){\strut{} 9}}%
      \put(4705,374){\makebox(0,0){\strut{} 10}}%
      \put(220,1864){\rotatebox{-270}{\makebox(0,0){\strut{}time (ns)}}}%
      \put(2715,154){\makebox(0,0){\strut{}E (MV/m)}}%
    }%
    \gplgaddtomacro\gplfronttext{%
      \csname LTb\endcsname%
      \put(3982,2962){\makebox(0,0)[r]{\strut{}fluid $(10^{11} \, \textrm{m}^{-3})$}}%
      \csname LTb\endcsname%
      \put(3982,2742){\makebox(0,0)[r]{\strut{}part $(10^{11} \, \textrm{m}^{-3})$}}%
      \csname LTb\endcsname%
      \put(3982,2522){\makebox(0,0)[r]{\strut{}$\tau_\mathrm{is}$ $(10^{11} \, \textrm{m}^{-3})$}}%
      \csname LTb\endcsname%
      \put(3982,2302){\makebox(0,0)[r]{\strut{}fluid $(10^{13} \, \textrm{m}^{-3})$}}%
      \csname LTb\endcsname%
      \put(3982,2082){\makebox(0,0)[r]{\strut{}part $(10^{13} \, \textrm{m}^{-3})$}}%
      \csname LTb\endcsname%
      \put(3982,1862){\makebox(0,0)[r]{\strut{}$\tau_\mathrm{is}$ $(10^{13} \, \textrm{m}^{-3})$}}%
    }%
    \gplbacktext
    \put(0,0){\includegraphics{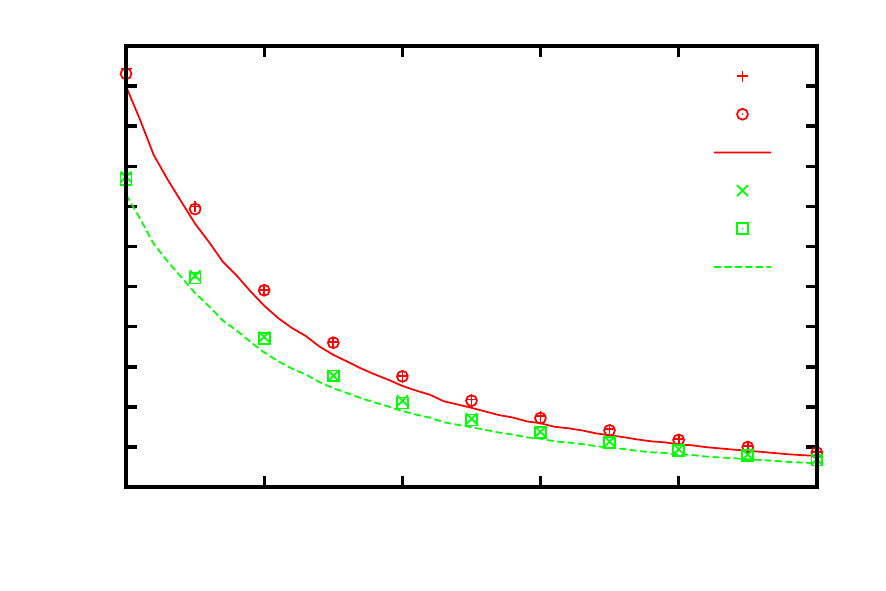}}%
    \gplfronttext
  \end{picture}%
\endgroup

%% file: fig-partial-screening-field_pdf.tex
\begingroup
  \makeatletter
  \providecommand\color[2][]{%
    \GenericError{(gnuplot) \space\space\space\@spaces}{%
      Package color not loaded in conjunction with
      terminal option `colourtext'%
    }{See the gnuplot documentation for explanation.%
    }{Either use 'blacktext' in gnuplot or load the package
      color.sty in LaTeX.}%
    \renewcommand\color[2][]{}%
  }%
  \providecommand\includegraphics[2][]{%
    \GenericError{(gnuplot) \space\space\space\@spaces}{%
      Package graphicx or graphics not loaded%
    }{See the gnuplot documentation for explanation.%
    }{The gnuplot epslatex terminal needs graphicx.sty or graphics.sty.}%
    \renewcommand\includegraphics[2][]{}%
  }%
  \providecommand\rotatebox[2]{#2}%
  \@ifundefined{ifGPcolor}{%
    \newif\ifGPcolor
    \GPcolortrue
  }{}%
  \@ifundefined{ifGPblacktext}{%
    \newif\ifGPblacktext
    \GPblacktexttrue
  }{}%
  \let\gplgaddtomacro\g@addto@macro
  \gdef\gplbacktext{}%
  \gdef\gplfronttext{}%
  \makeatother
  \ifGPblacktext
    \def\colorrgb#1{}%
    \def\colorgray#1{}%
  \else
    \ifGPcolor
      \def\colorrgb#1{\color[rgb]{#1}}%
      \def\colorgray#1{\color[gray]{#1}}%
      \expandafter\def\csname LTw\endcsname{\color{white}}%
      \expandafter\def\csname LTb\endcsname{\color{black}}%
      \expandafter\def\csname LTa\endcsname{\color{black}}%
      \expandafter\def\csname LT0\endcsname{\color[rgb]{1,0,0}}%
      \expandafter\def\csname LT1\endcsname{\color[rgb]{0,1,0}}%
      \expandafter\def\csname LT2\endcsname{\color[rgb]{0,0,1}}%
      \expandafter\def\csname LT3\endcsname{\color[rgb]{1,0,1}}%
      \expandafter\def\csname LT4\endcsname{\color[rgb]{0,1,1}}%
      \expandafter\def\csname LT5\endcsname{\color[rgb]{1,1,0}}%
      \expandafter\def\csname LT6\endcsname{\color[rgb]{0,0,0}}%
      \expandafter\def\csname LT7\endcsname{\color[rgb]{1,0.3,0}}%
      \expandafter\def\csname LT8\endcsname{\color[rgb]{0.5,0.5,0.5}}%
    \else
      \def\colorrgb#1{\color{black}}%
      \def\colorgray#1{\color[gray]{#1}}%
      \expandafter\def\csname LTw\endcsname{\color{white}}%
      \expandafter\def\csname LTb\endcsname{\color{black}}%
      \expandafter\def\csname LTa\endcsname{\color{black}}%
      \expandafter\def\csname LT0\endcsname{\color{black}}%
      \expandafter\def\csname LT1\endcsname{\color{black}}%
      \expandafter\def\csname LT2\endcsname{\color{black}}%
      \expandafter\def\csname LT3\endcsname{\color{black}}%
      \expandafter\def\csname LT4\endcsname{\color{black}}%
      \expandafter\def\csname LT5\endcsname{\color{black}}%
      \expandafter\def\csname LT6\endcsname{\color{black}}%
      \expandafter\def\csname LT7\endcsname{\color{black}}%
      \expandafter\def\csname LT8\endcsname{\color{black}}%
    \fi
  \fi
  \setlength{\unitlength}{0.0500bp}%
  \begin{picture}(4534.00,3174.00)%
    \gplgaddtomacro\gplbacktext{%
      \csname LTb\endcsname%
      \put(726,594){\makebox(0,0)[r]{\strut{} 2.5}}%
      \put(726,883){\makebox(0,0)[r]{\strut{} 3}}%
      \put(726,1173){\makebox(0,0)[r]{\strut{} 3.5}}%
      \put(726,1462){\makebox(0,0)[r]{\strut{} 4}}%
      \put(726,1752){\makebox(0,0)[r]{\strut{} 4.5}}%
      \put(726,2041){\makebox(0,0)[r]{\strut{} 5}}%
      \put(726,2330){\makebox(0,0)[r]{\strut{} 5.5}}%
      \put(726,2620){\makebox(0,0)[r]{\strut{} 6}}%
      \put(726,2909){\makebox(0,0)[r]{\strut{} 6.5}}%
      \put(1156,374){\makebox(0,0){\strut{} 10}}%
      \put(1752,374){\makebox(0,0){\strut{} 11}}%
      \put(2348,374){\makebox(0,0){\strut{} 12}}%
      \put(2945,374){\makebox(0,0){\strut{} 13}}%
      \put(3541,374){\makebox(0,0){\strut{} 14}}%
      \put(4137,374){\makebox(0,0){\strut{} 15}}%
      \put(220,1751){\rotatebox{-270}{\makebox(0,0){\strut{}E (MV/m)}}}%
      \put(2497,154){\makebox(0,0){\strut{}x (mm)}}%
    }%
    \gplgaddtomacro\gplfronttext{%
      \csname LTb\endcsname%
      \put(3414,2087){\makebox(0,0)[r]{\strut{}2.1 ns}}%
      \csname LTb\endcsname%
      \put(3414,1867){\makebox(0,0)[r]{\strut{}2.4 ns}}%
      \csname LTb\endcsname%
      \put(3414,1647){\makebox(0,0)[r]{\strut{}2.7 ns}}%
      \csname LTb\endcsname%
      \put(3414,1427){\makebox(0,0)[r]{\strut{}3.0 ns}}%
      \csname LTb\endcsname%
      \put(3414,1207){\makebox(0,0)[r]{\strut{}3.3 ns}}%
      \csname LTb\endcsname%
      \put(3414,987){\makebox(0,0)[r]{\strut{}3.6 ns}}%
      \csname LTb\endcsname%
      \put(3414,767){\makebox(0,0)[r]{\strut{}3.9 ns}}%
    }%
    \gplbacktext
    \put(0,0){\includegraphics{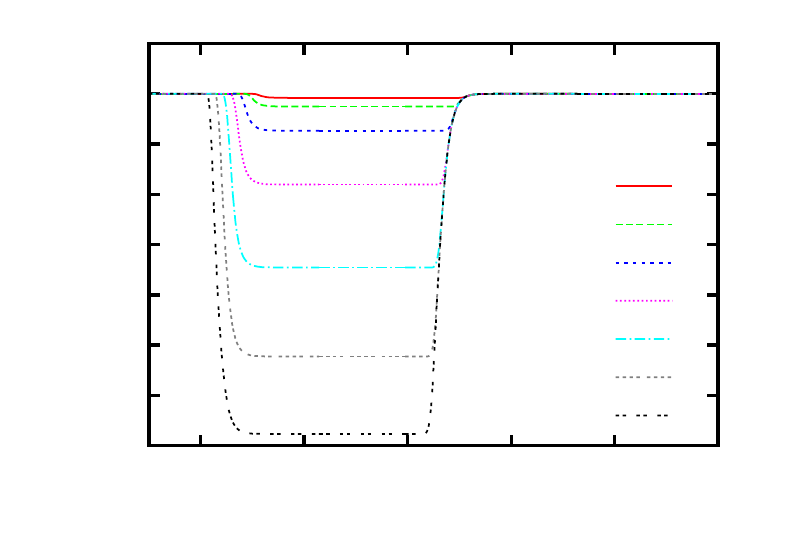}}%
    \gplfronttext
  \end{picture}%
\endgroup

%% file: fig-partial-screening-dens_pdf.tex
\begingroup
  \makeatletter
  \providecommand\color[2][]{%
    \GenericError{(gnuplot) \space\space\space\@spaces}{%
      Package color not loaded in conjunction with
      terminal option `colourtext'%
    }{See the gnuplot documentation for explanation.%
    }{Either use 'blacktext' in gnuplot or load the package
      color.sty in LaTeX.}%
    \renewcommand\color[2][]{}%
  }%
  \providecommand\includegraphics[2][]{%
    \GenericError{(gnuplot) \space\space\space\@spaces}{%
      Package graphicx or graphics not loaded%
    }{See the gnuplot documentation for explanation.%
    }{The gnuplot epslatex terminal needs graphicx.sty or graphics.sty.}%
    \renewcommand\includegraphics[2][]{}%
  }%
  \providecommand\rotatebox[2]{#2}%
  \@ifundefined{ifGPcolor}{%
    \newif\ifGPcolor
    \GPcolortrue
  }{}%
  \@ifundefined{ifGPblacktext}{%
    \newif\ifGPblacktext
    \GPblacktexttrue
  }{}%
  \let\gplgaddtomacro\g@addto@macro
  \gdef\gplbacktext{}%
  \gdef\gplfronttext{}%
  \makeatother
  \ifGPblacktext
    \def\colorrgb#1{}%
    \def\colorgray#1{}%
  \else
    \ifGPcolor
      \def\colorrgb#1{\color[rgb]{#1}}%
      \def\colorgray#1{\color[gray]{#1}}%
      \expandafter\def\csname LTw\endcsname{\color{white}}%
      \expandafter\def\csname LTb\endcsname{\color{black}}%
      \expandafter\def\csname LTa\endcsname{\color{black}}%
      \expandafter\def\csname LT0\endcsname{\color[rgb]{1,0,0}}%
      \expandafter\def\csname LT1\endcsname{\color[rgb]{0,1,0}}%
      \expandafter\def\csname LT2\endcsname{\color[rgb]{0,0,1}}%
      \expandafter\def\csname LT3\endcsname{\color[rgb]{1,0,1}}%
      \expandafter\def\csname LT4\endcsname{\color[rgb]{0,1,1}}%
      \expandafter\def\csname LT5\endcsname{\color[rgb]{1,1,0}}%
      \expandafter\def\csname LT6\endcsname{\color[rgb]{0,0,0}}%
      \expandafter\def\csname LT7\endcsname{\color[rgb]{1,0.3,0}}%
      \expandafter\def\csname LT8\endcsname{\color[rgb]{0.5,0.5,0.5}}%
    \else
      \def\colorrgb#1{\color{black}}%
      \def\colorgray#1{\color[gray]{#1}}%
      \expandafter\def\csname LTw\endcsname{\color{white}}%
      \expandafter\def\csname LTb\endcsname{\color{black}}%
      \expandafter\def\csname LTa\endcsname{\color{black}}%
      \expandafter\def\csname LT0\endcsname{\color{black}}%
      \expandafter\def\csname LT1\endcsname{\color{black}}%
      \expandafter\def\csname LT2\endcsname{\color{black}}%
      \expandafter\def\csname LT3\endcsname{\color{black}}%
      \expandafter\def\csname LT4\endcsname{\color{black}}%
      \expandafter\def\csname LT5\endcsname{\color{black}}%
      \expandafter\def\csname LT6\endcsname{\color{black}}%
      \expandafter\def\csname LT7\endcsname{\color{black}}%
      \expandafter\def\csname LT8\endcsname{\color{black}}%
    \fi
  \fi
  \setlength{\unitlength}{0.0500bp}%
  \begin{picture}(4534.00,3174.00)%
    \gplgaddtomacro\gplbacktext{%
      \csname LTb\endcsname%
      \put(726,1224){\makebox(0,0)[r]{\strut{} 0.1}}%
      \put(726,2429){\makebox(0,0)[r]{\strut{} 1}}%
      \put(1156,374){\makebox(0,0){\strut{} 10}}%
      \put(1752,374){\makebox(0,0){\strut{} 11}}%
      \put(2348,374){\makebox(0,0){\strut{} 12}}%
      \put(2945,374){\makebox(0,0){\strut{} 13}}%
      \put(3541,374){\makebox(0,0){\strut{} 14}}%
      \put(4137,374){\makebox(0,0){\strut{} 15}}%
      \put(220,1751){\rotatebox{-270}{\makebox(0,0){\strut{}$n_e$ ($10^{18}\,\mathrm{m}^{-3}$)}}}%
      \put(2497,154){\makebox(0,0){\strut{}x (mm)}}%
    }%
    \gplgaddtomacro\gplfronttext{%
      \csname LTb\endcsname%
      \put(3414,2087){\makebox(0,0)[r]{\strut{}2.1 ns}}%
      \csname LTb\endcsname%
      \put(3414,1867){\makebox(0,0)[r]{\strut{}2.4 ns}}%
      \csname LTb\endcsname%
      \put(3414,1647){\makebox(0,0)[r]{\strut{}2.7 ns}}%
      \csname LTb\endcsname%
      \put(3414,1427){\makebox(0,0)[r]{\strut{}3.0 ns}}%
      \csname LTb\endcsname%
      \put(3414,1207){\makebox(0,0)[r]{\strut{}3.3 ns}}%
      \csname LTb\endcsname%
      \put(3414,987){\makebox(0,0)[r]{\strut{}3.6 ns}}%
      \csname LTb\endcsname%
      \put(3414,767){\makebox(0,0)[r]{\strut{}3.9 ns}}%
    }%
    \gplbacktext
    \put(0,0){\includegraphics{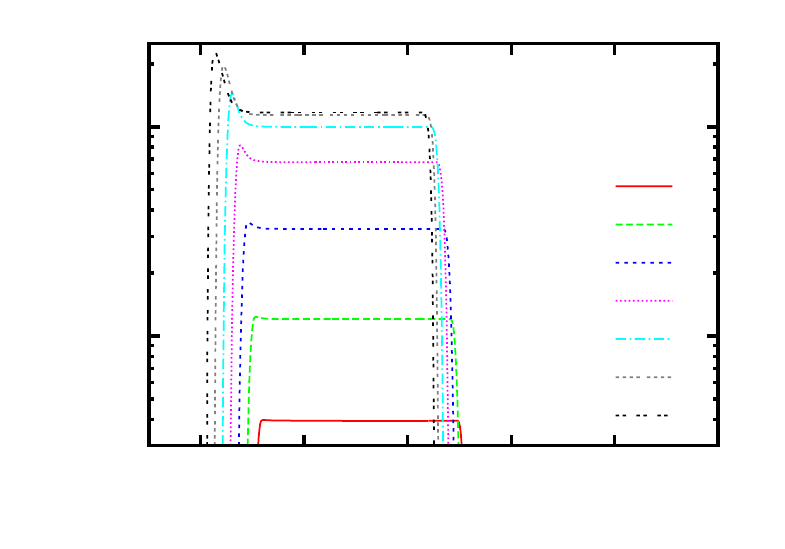}}%
    \gplfronttext
  \end{picture}%
\endgroup

%% file: fig-3d-screening-1e13-field_pdf.tex
\begingroup
  \makeatletter
  \providecommand\color[2][]{%
    \GenericError{(gnuplot) \space\space\space\@spaces}{%
      Package color not loaded in conjunction with
      terminal option `colourtext'%
    }{See the gnuplot documentation for explanation.%
    }{Either use 'blacktext' in gnuplot or load the package
      color.sty in LaTeX.}%
    \renewcommand\color[2][]{}%
  }%
  \providecommand\includegraphics[2][]{%
    \GenericError{(gnuplot) \space\space\space\@spaces}{%
      Package graphicx or graphics not loaded%
    }{See the gnuplot documentation for explanation.%
    }{The gnuplot epslatex terminal needs graphicx.sty or graphics.sty.}%
    \renewcommand\includegraphics[2][]{}%
  }%
  \providecommand\rotatebox[2]{#2}%
  \@ifundefined{ifGPcolor}{%
    \newif\ifGPcolor
    \GPcolortrue
  }{}%
  \@ifundefined{ifGPblacktext}{%
    \newif\ifGPblacktext
    \GPblacktexttrue
  }{}%
  \let\gplgaddtomacro\g@addto@macro
  \gdef\gplbacktext{}%
  \gdef\gplfronttext{}%
  \makeatother
  \ifGPblacktext
    \def\colorrgb#1{}%
    \def\colorgray#1{}%
  \else
    \ifGPcolor
      \def\colorrgb#1{\color[rgb]{#1}}%
      \def\colorgray#1{\color[gray]{#1}}%
      \expandafter\def\csname LTw\endcsname{\color{white}}%
      \expandafter\def\csname LTb\endcsname{\color{black}}%
      \expandafter\def\csname LTa\endcsname{\color{black}}%
      \expandafter\def\csname LT0\endcsname{\color[rgb]{1,0,0}}%
      \expandafter\def\csname LT1\endcsname{\color[rgb]{0,1,0}}%
      \expandafter\def\csname LT2\endcsname{\color[rgb]{0,0,1}}%
      \expandafter\def\csname LT3\endcsname{\color[rgb]{1,0,1}}%
      \expandafter\def\csname LT4\endcsname{\color[rgb]{0,1,1}}%
      \expandafter\def\csname LT5\endcsname{\color[rgb]{1,1,0}}%
      \expandafter\def\csname LT6\endcsname{\color[rgb]{0,0,0}}%
      \expandafter\def\csname LT7\endcsname{\color[rgb]{1,0.3,0}}%
      \expandafter\def\csname LT8\endcsname{\color[rgb]{0.5,0.5,0.5}}%
    \else
      \def\colorrgb#1{\color{black}}%
      \def\colorgray#1{\color[gray]{#1}}%
      \expandafter\def\csname LTw\endcsname{\color{white}}%
      \expandafter\def\csname LTb\endcsname{\color{black}}%
      \expandafter\def\csname LTa\endcsname{\color{black}}%
      \expandafter\def\csname LT0\endcsname{\color{black}}%
      \expandafter\def\csname LT1\endcsname{\color{black}}%
      \expandafter\def\csname LT2\endcsname{\color{black}}%
      \expandafter\def\csname LT3\endcsname{\color{black}}%
      \expandafter\def\csname LT4\endcsname{\color{black}}%
      \expandafter\def\csname LT5\endcsname{\color{black}}%
      \expandafter\def\csname LT6\endcsname{\color{black}}%
      \expandafter\def\csname LT7\endcsname{\color{black}}%
      \expandafter\def\csname LT8\endcsname{\color{black}}%
    \fi
  \fi
  \setlength{\unitlength}{0.0500bp}%
  \begin{picture}(4534.00,3174.00)%
    \gplgaddtomacro\gplbacktext{%
      \csname LTb\endcsname%
      \put(726,594){\makebox(0,0)[r]{\strut{} 2}}%
      \put(726,851){\makebox(0,0)[r]{\strut{} 2.5}}%
      \put(726,1108){\makebox(0,0)[r]{\strut{} 3}}%
      \put(726,1366){\makebox(0,0)[r]{\strut{} 3.5}}%
      \put(726,1623){\makebox(0,0)[r]{\strut{} 4}}%
      \put(726,1880){\makebox(0,0)[r]{\strut{} 4.5}}%
      \put(726,2137){\makebox(0,0)[r]{\strut{} 5}}%
      \put(726,2395){\makebox(0,0)[r]{\strut{} 5.5}}%
      \put(726,2652){\makebox(0,0)[r]{\strut{} 6}}%
      \put(726,2909){\makebox(0,0)[r]{\strut{} 6.5}}%
      \put(1286,374){\makebox(0,0){\strut{} 3}}%
      \put(1856,374){\makebox(0,0){\strut{} 4}}%
      \put(2426,374){\makebox(0,0){\strut{} 5}}%
      \put(2996,374){\makebox(0,0){\strut{} 6}}%
      \put(3567,374){\makebox(0,0){\strut{} 7}}%
      \put(4137,374){\makebox(0,0){\strut{} 8}}%
      \put(220,1751){\rotatebox{-270}{\makebox(0,0){\strut{}E (MV/m)}}}%
      \put(2497,154){\makebox(0,0){\strut{}x (mm)}}%
    }%
    \gplgaddtomacro\gplfronttext{%
      \csname LTb\endcsname%
      \put(3414,1867){\makebox(0,0)[r]{\strut{}2.55 ns}}%
      \csname LTb\endcsname%
      \put(3414,1647){\makebox(0,0)[r]{\strut{}2.85 ns}}%
      \csname LTb\endcsname%
      \put(3414,1427){\makebox(0,0)[r]{\strut{}3.15 ns}}%
      \csname LTb\endcsname%
      \put(3414,1207){\makebox(0,0)[r]{\strut{}3.45 ns}}%
      \csname LTb\endcsname%
      \put(3414,987){\makebox(0,0)[r]{\strut{}3.75 ns}}%
      \csname LTb\endcsname%
      \put(3414,767){\makebox(0,0)[r]{\strut{}4.05 ns}}%
    }%
    \gplbacktext
    \put(0,0){\includegraphics{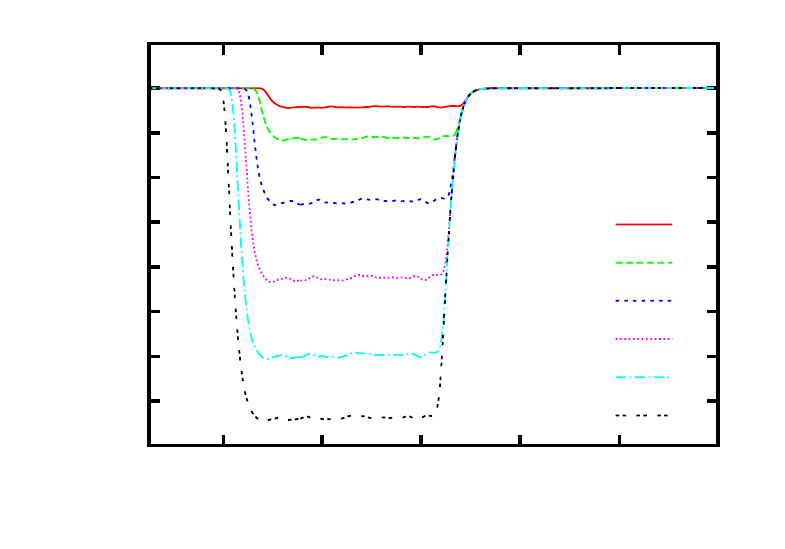}}%
    \gplfronttext
  \end{picture}%
\endgroup

%% file: fig-3d-screening-1e13-dens_pdf.tex
\begingroup
  \makeatletter
  \providecommand\color[2][]{%
    \GenericError{(gnuplot) \space\space\space\@spaces}{%
      Package color not loaded in conjunction with
      terminal option `colourtext'%
    }{See the gnuplot documentation for explanation.%
    }{Either use 'blacktext' in gnuplot or load the package
      color.sty in LaTeX.}%
    \renewcommand\color[2][]{}%
  }%
  \providecommand\includegraphics[2][]{%
    \GenericError{(gnuplot) \space\space\space\@spaces}{%
      Package graphicx or graphics not loaded%
    }{See the gnuplot documentation for explanation.%
    }{The gnuplot epslatex terminal needs graphicx.sty or graphics.sty.}%
    \renewcommand\includegraphics[2][]{}%
  }%
  \providecommand\rotatebox[2]{#2}%
  \@ifundefined{ifGPcolor}{%
    \newif\ifGPcolor
    \GPcolortrue
  }{}%
  \@ifundefined{ifGPblacktext}{%
    \newif\ifGPblacktext
    \GPblacktexttrue
  }{}%
  \let\gplgaddtomacro\g@addto@macro
  \gdef\gplbacktext{}%
  \gdef\gplfronttext{}%
  \makeatother
  \ifGPblacktext
    \def\colorrgb#1{}%
    \def\colorgray#1{}%
  \else
    \ifGPcolor
      \def\colorrgb#1{\color[rgb]{#1}}%
      \def\colorgray#1{\color[gray]{#1}}%
      \expandafter\def\csname LTw\endcsname{\color{white}}%
      \expandafter\def\csname LTb\endcsname{\color{black}}%
      \expandafter\def\csname LTa\endcsname{\color{black}}%
      \expandafter\def\csname LT0\endcsname{\color[rgb]{1,0,0}}%
      \expandafter\def\csname LT1\endcsname{\color[rgb]{0,1,0}}%
      \expandafter\def\csname LT2\endcsname{\color[rgb]{0,0,1}}%
      \expandafter\def\csname LT3\endcsname{\color[rgb]{1,0,1}}%
      \expandafter\def\csname LT4\endcsname{\color[rgb]{0,1,1}}%
      \expandafter\def\csname LT5\endcsname{\color[rgb]{1,1,0}}%
      \expandafter\def\csname LT6\endcsname{\color[rgb]{0,0,0}}%
      \expandafter\def\csname LT7\endcsname{\color[rgb]{1,0.3,0}}%
      \expandafter\def\csname LT8\endcsname{\color[rgb]{0.5,0.5,0.5}}%
    \else
      \def\colorrgb#1{\color{black}}%
      \def\colorgray#1{\color[gray]{#1}}%
      \expandafter\def\csname LTw\endcsname{\color{white}}%
      \expandafter\def\csname LTb\endcsname{\color{black}}%
      \expandafter\def\csname LTa\endcsname{\color{black}}%
      \expandafter\def\csname LT0\endcsname{\color{black}}%
      \expandafter\def\csname LT1\endcsname{\color{black}}%
      \expandafter\def\csname LT2\endcsname{\color{black}}%
      \expandafter\def\csname LT3\endcsname{\color{black}}%
      \expandafter\def\csname LT4\endcsname{\color{black}}%
      \expandafter\def\csname LT5\endcsname{\color{black}}%
      \expandafter\def\csname LT6\endcsname{\color{black}}%
      \expandafter\def\csname LT7\endcsname{\color{black}}%
      \expandafter\def\csname LT8\endcsname{\color{black}}%
    \fi
  \fi
  \setlength{\unitlength}{0.0500bp}%
  \begin{picture}(4534.00,3174.00)%
    \gplgaddtomacro\gplbacktext{%
      \csname LTb\endcsname%
      \put(726,986){\makebox(0,0)[r]{\strut{} 0.1}}%
      \put(726,2288){\makebox(0,0)[r]{\strut{} 1}}%
      \put(1286,374){\makebox(0,0){\strut{} 3}}%
      \put(1856,374){\makebox(0,0){\strut{} 4}}%
      \put(2426,374){\makebox(0,0){\strut{} 5}}%
      \put(2996,374){\makebox(0,0){\strut{} 6}}%
      \put(3567,374){\makebox(0,0){\strut{} 7}}%
      \put(4137,374){\makebox(0,0){\strut{} 8}}%
      \put(220,1751){\rotatebox{-270}{\makebox(0,0){\strut{}$n_e$ ($10^{18}\,\mathrm{m}^{-3}$)}}}%
      \put(2497,154){\makebox(0,0){\strut{}x (mm)}}%
    }%
    \gplgaddtomacro\gplfronttext{%
      \csname LTb\endcsname%
      \put(3414,1867){\makebox(0,0)[r]{\strut{}2.55 ns}}%
      \csname LTb\endcsname%
      \put(3414,1647){\makebox(0,0)[r]{\strut{}2.85 ns}}%
      \csname LTb\endcsname%
      \put(3414,1427){\makebox(0,0)[r]{\strut{}3.15 ns}}%
      \csname LTb\endcsname%
      \put(3414,1207){\makebox(0,0)[r]{\strut{}3.45 ns}}%
      \csname LTb\endcsname%
      \put(3414,987){\makebox(0,0)[r]{\strut{}3.75 ns}}%
      \csname LTb\endcsname%
      \put(3414,767){\makebox(0,0)[r]{\strut{}4.05 ns}}%
    }%
    \gplbacktext
    \put(0,0){\includegraphics{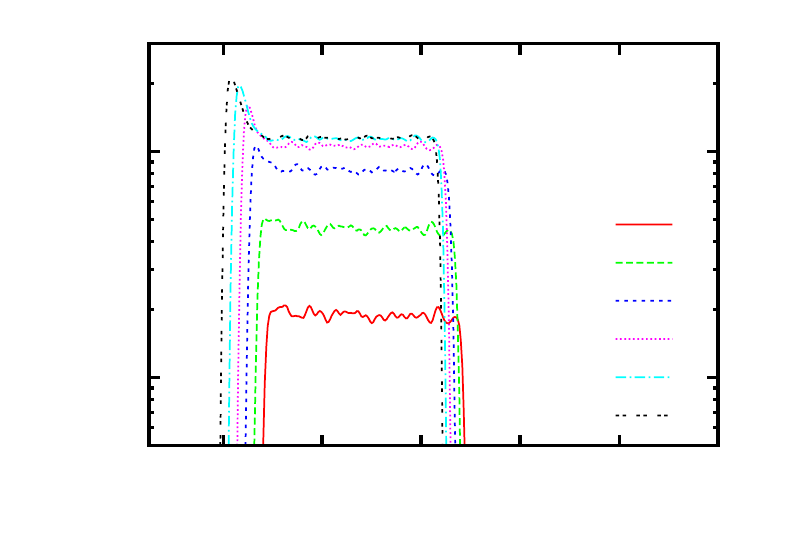}}%
    \gplfronttext
  \end{picture}%
\endgroup

%% file: fig-3d-screening-1e11-field_pdf.tex
\begingroup
  \makeatletter
  \providecommand\color[2][]{%
    \GenericError{(gnuplot) \space\space\space\@spaces}{%
      Package color not loaded in conjunction with
      terminal option `colourtext'%
    }{See the gnuplot documentation for explanation.%
    }{Either use 'blacktext' in gnuplot or load the package
      color.sty in LaTeX.}%
    \renewcommand\color[2][]{}%
  }%
  \providecommand\includegraphics[2][]{%
    \GenericError{(gnuplot) \space\space\space\@spaces}{%
      Package graphicx or graphics not loaded%
    }{See the gnuplot documentation for explanation.%
    }{The gnuplot epslatex terminal needs graphicx.sty or graphics.sty.}%
    \renewcommand\includegraphics[2][]{}%
  }%
  \providecommand\rotatebox[2]{#2}%
  \@ifundefined{ifGPcolor}{%
    \newif\ifGPcolor
    \GPcolortrue
  }{}%
  \@ifundefined{ifGPblacktext}{%
    \newif\ifGPblacktext
    \GPblacktexttrue
  }{}%
  \let\gplgaddtomacro\g@addto@macro
  \gdef\gplbacktext{}%
  \gdef\gplfronttext{}%
  \makeatother
  \ifGPblacktext
    \def\colorrgb#1{}%
    \def\colorgray#1{}%
  \else
    \ifGPcolor
      \def\colorrgb#1{\color[rgb]{#1}}%
      \def\colorgray#1{\color[gray]{#1}}%
      \expandafter\def\csname LTw\endcsname{\color{white}}%
      \expandafter\def\csname LTb\endcsname{\color{black}}%
      \expandafter\def\csname LTa\endcsname{\color{black}}%
      \expandafter\def\csname LT0\endcsname{\color[rgb]{1,0,0}}%
      \expandafter\def\csname LT1\endcsname{\color[rgb]{0,1,0}}%
      \expandafter\def\csname LT2\endcsname{\color[rgb]{0,0,1}}%
      \expandafter\def\csname LT3\endcsname{\color[rgb]{1,0,1}}%
      \expandafter\def\csname LT4\endcsname{\color[rgb]{0,1,1}}%
      \expandafter\def\csname LT5\endcsname{\color[rgb]{1,1,0}}%
      \expandafter\def\csname LT6\endcsname{\color[rgb]{0,0,0}}%
      \expandafter\def\csname LT7\endcsname{\color[rgb]{1,0.3,0}}%
      \expandafter\def\csname LT8\endcsname{\color[rgb]{0.5,0.5,0.5}}%
    \else
      \def\colorrgb#1{\color{black}}%
      \def\colorgray#1{\color[gray]{#1}}%
      \expandafter\def\csname LTw\endcsname{\color{white}}%
      \expandafter\def\csname LTb\endcsname{\color{black}}%
      \expandafter\def\csname LTa\endcsname{\color{black}}%
      \expandafter\def\csname LT0\endcsname{\color{black}}%
      \expandafter\def\csname LT1\endcsname{\color{black}}%
      \expandafter\def\csname LT2\endcsname{\color{black}}%
      \expandafter\def\csname LT3\endcsname{\color{black}}%
      \expandafter\def\csname LT4\endcsname{\color{black}}%
      \expandafter\def\csname LT5\endcsname{\color{black}}%
      \expandafter\def\csname LT6\endcsname{\color{black}}%
      \expandafter\def\csname LT7\endcsname{\color{black}}%
      \expandafter\def\csname LT8\endcsname{\color{black}}%
    \fi
  \fi
  \setlength{\unitlength}{0.0500bp}%
  \begin{picture}(4534.00,3174.00)%
    \gplgaddtomacro\gplbacktext{%
      \csname LTb\endcsname%
      \put(726,594){\makebox(0,0)[r]{\strut{} 2}}%
      \put(726,851){\makebox(0,0)[r]{\strut{} 2.5}}%
      \put(726,1108){\makebox(0,0)[r]{\strut{} 3}}%
      \put(726,1366){\makebox(0,0)[r]{\strut{} 3.5}}%
      \put(726,1623){\makebox(0,0)[r]{\strut{} 4}}%
      \put(726,1880){\makebox(0,0)[r]{\strut{} 4.5}}%
      \put(726,2137){\makebox(0,0)[r]{\strut{} 5}}%
      \put(726,2395){\makebox(0,0)[r]{\strut{} 5.5}}%
      \put(726,2652){\makebox(0,0)[r]{\strut{} 6}}%
      \put(726,2909){\makebox(0,0)[r]{\strut{} 6.5}}%
      \put(1286,374){\makebox(0,0){\strut{} 3}}%
      \put(1856,374){\makebox(0,0){\strut{} 4}}%
      \put(2426,374){\makebox(0,0){\strut{} 5}}%
      \put(2996,374){\makebox(0,0){\strut{} 6}}%
      \put(3567,374){\makebox(0,0){\strut{} 7}}%
      \put(4137,374){\makebox(0,0){\strut{} 8}}%
      \put(220,1751){\rotatebox{-270}{\makebox(0,0){\strut{}E (MV/m)}}}%
      \put(2497,154){\makebox(0,0){\strut{}x (mm)}}%
    }%
    \gplgaddtomacro\gplfronttext{%
      \csname LTb\endcsname%
      \put(3414,1867){\makebox(0,0)[r]{\strut{}3.75 ns}}%
      \csname LTb\endcsname%
      \put(3414,1647){\makebox(0,0)[r]{\strut{}4.05 ns}}%
      \csname LTb\endcsname%
      \put(3414,1427){\makebox(0,0)[r]{\strut{}4.35 ns}}%
      \csname LTb\endcsname%
      \put(3414,1207){\makebox(0,0)[r]{\strut{}4.65 ns}}%
      \csname LTb\endcsname%
      \put(3414,987){\makebox(0,0)[r]{\strut{}4.95 ns}}%
      \csname LTb\endcsname%
      \put(3414,767){\makebox(0,0)[r]{\strut{}5.25 ns}}%
    }%
    \gplbacktext
    \put(0,0){\includegraphics{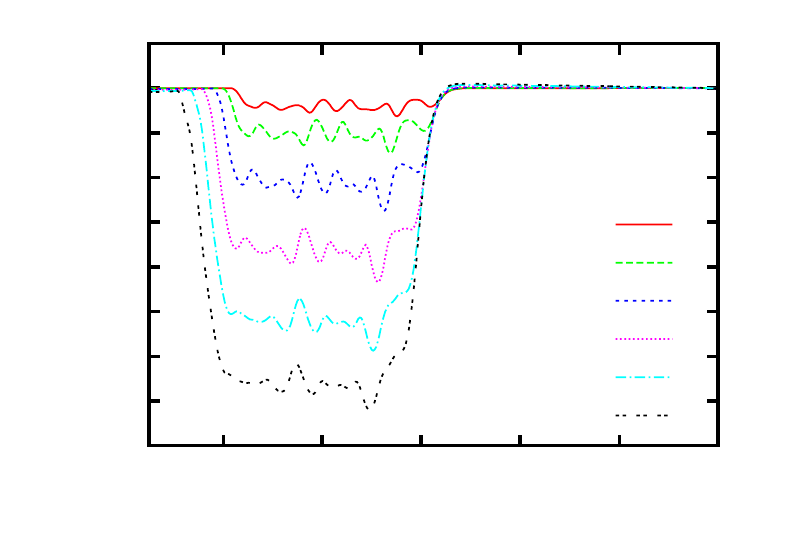}}%
    \gplfronttext
  \end{picture}%
\endgroup

%% file: fig-3d-screening-1e11-dens_pdf.tex
\begingroup
  \makeatletter
  \providecommand\color[2][]{%
    \GenericError{(gnuplot) \space\space\space\@spaces}{%
      Package color not loaded in conjunction with
      terminal option `colourtext'%
    }{See the gnuplot documentation for explanation.%
    }{Either use 'blacktext' in gnuplot or load the package
      color.sty in LaTeX.}%
    \renewcommand\color[2][]{}%
  }%
  \providecommand\includegraphics[2][]{%
    \GenericError{(gnuplot) \space\space\space\@spaces}{%
      Package graphicx or graphics not loaded%
    }{See the gnuplot documentation for explanation.%
    }{The gnuplot epslatex terminal needs graphicx.sty or graphics.sty.}%
    \renewcommand\includegraphics[2][]{}%
  }%
  \providecommand\rotatebox[2]{#2}%
  \@ifundefined{ifGPcolor}{%
    \newif\ifGPcolor
    \GPcolortrue
  }{}%
  \@ifundefined{ifGPblacktext}{%
    \newif\ifGPblacktext
    \GPblacktexttrue
  }{}%
  \let\gplgaddtomacro\g@addto@macro
  \gdef\gplbacktext{}%
  \gdef\gplfronttext{}%
  \makeatother
  \ifGPblacktext
    \def\colorrgb#1{}%
    \def\colorgray#1{}%
  \else
    \ifGPcolor
      \def\colorrgb#1{\color[rgb]{#1}}%
      \def\colorgray#1{\color[gray]{#1}}%
      \expandafter\def\csname LTw\endcsname{\color{white}}%
      \expandafter\def\csname LTb\endcsname{\color{black}}%
      \expandafter\def\csname LTa\endcsname{\color{black}}%
      \expandafter\def\csname LT0\endcsname{\color[rgb]{1,0,0}}%
      \expandafter\def\csname LT1\endcsname{\color[rgb]{0,1,0}}%
      \expandafter\def\csname LT2\endcsname{\color[rgb]{0,0,1}}%
      \expandafter\def\csname LT3\endcsname{\color[rgb]{1,0,1}}%
      \expandafter\def\csname LT4\endcsname{\color[rgb]{0,1,1}}%
      \expandafter\def\csname LT5\endcsname{\color[rgb]{1,1,0}}%
      \expandafter\def\csname LT6\endcsname{\color[rgb]{0,0,0}}%
      \expandafter\def\csname LT7\endcsname{\color[rgb]{1,0.3,0}}%
      \expandafter\def\csname LT8\endcsname{\color[rgb]{0.5,0.5,0.5}}%
    \else
      \def\colorrgb#1{\color{black}}%
      \def\colorgray#1{\color[gray]{#1}}%
      \expandafter\def\csname LTw\endcsname{\color{white}}%
      \expandafter\def\csname LTb\endcsname{\color{black}}%
      \expandafter\def\csname LTa\endcsname{\color{black}}%
      \expandafter\def\csname LT0\endcsname{\color{black}}%
      \expandafter\def\csname LT1\endcsname{\color{black}}%
      \expandafter\def\csname LT2\endcsname{\color{black}}%
      \expandafter\def\csname LT3\endcsname{\color{black}}%
      \expandafter\def\csname LT4\endcsname{\color{black}}%
      \expandafter\def\csname LT5\endcsname{\color{black}}%
      \expandafter\def\csname LT6\endcsname{\color{black}}%
      \expandafter\def\csname LT7\endcsname{\color{black}}%
      \expandafter\def\csname LT8\endcsname{\color{black}}%
    \fi
  \fi
  \setlength{\unitlength}{0.0500bp}%
  \begin{picture}(4534.00,3174.00)%
    \gplgaddtomacro\gplbacktext{%
      \csname LTb\endcsname%
      \put(726,986){\makebox(0,0)[r]{\strut{} 0.1}}%
      \put(726,2288){\makebox(0,0)[r]{\strut{} 1}}%
      \put(1286,374){\makebox(0,0){\strut{} 3}}%
      \put(1856,374){\makebox(0,0){\strut{} 4}}%
      \put(2426,374){\makebox(0,0){\strut{} 5}}%
      \put(2996,374){\makebox(0,0){\strut{} 6}}%
      \put(3567,374){\makebox(0,0){\strut{} 7}}%
      \put(4137,374){\makebox(0,0){\strut{} 8}}%
      \put(220,1751){\rotatebox{-270}{\makebox(0,0){\strut{}$n_e$ ($10^{18}\,\mathrm{m}^{-3}$)}}}%
      \put(2497,154){\makebox(0,0){\strut{}x (mm)}}%
    }%
    \gplgaddtomacro\gplfronttext{%
      \csname LTb\endcsname%
      \put(3414,1867){\makebox(0,0)[r]{\strut{}3.75 ns}}%
      \csname LTb\endcsname%
      \put(3414,1647){\makebox(0,0)[r]{\strut{}4.05 ns}}%
      \csname LTb\endcsname%
      \put(3414,1427){\makebox(0,0)[r]{\strut{}4.35 ns}}%
      \csname LTb\endcsname%
      \put(3414,1207){\makebox(0,0)[r]{\strut{}4.65 ns}}%
      \csname LTb\endcsname%
      \put(3414,987){\makebox(0,0)[r]{\strut{}4.95 ns}}%
      \csname LTb\endcsname%
      \put(3414,767){\makebox(0,0)[r]{\strut{}5.25 ns}}%
    }%
    \gplbacktext
    \put(0,0){\includegraphics{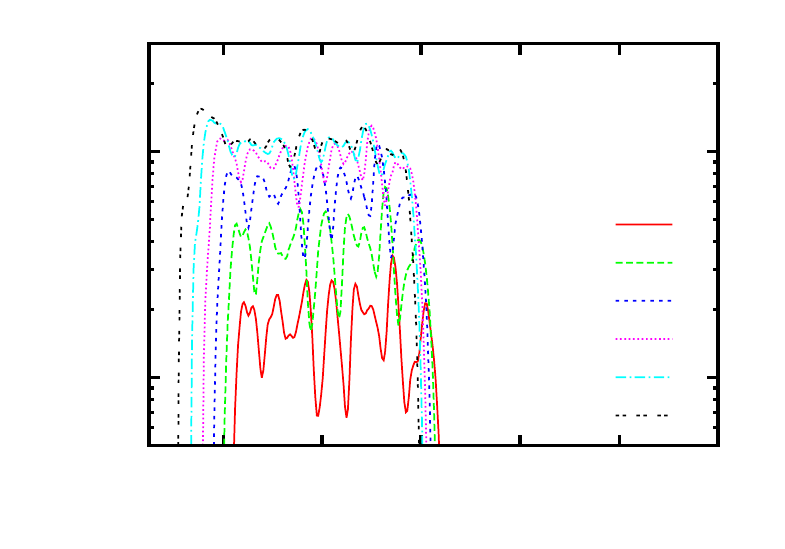}}%
    \gplfronttext
  \end{picture}%
\endgroup

%% file: fig-n0-min_pdf.tex
\begingroup
  \makeatletter
  \providecommand\color[2][]{%
    \GenericError{(gnuplot) \space\space\space\@spaces}{%
      Package color not loaded in conjunction with
      terminal option `colourtext'%
    }{See the gnuplot documentation for explanation.%
    }{Either use 'blacktext' in gnuplot or load the package
      color.sty in LaTeX.}%
    \renewcommand\color[2][]{}%
  }%
  \providecommand\includegraphics[2][]{%
    \GenericError{(gnuplot) \space\space\space\@spaces}{%
      Package graphicx or graphics not loaded%
    }{See the gnuplot documentation for explanation.%
    }{The gnuplot epslatex terminal needs graphicx.sty or graphics.sty.}%
    \renewcommand\includegraphics[2][]{}%
  }%
  \providecommand\rotatebox[2]{#2}%
  \@ifundefined{ifGPcolor}{%
    \newif\ifGPcolor
    \GPcolortrue
  }{}%
  \@ifundefined{ifGPblacktext}{%
    \newif\ifGPblacktext
    \GPblacktexttrue
  }{}%
  \let\gplgaddtomacro\g@addto@macro
  \gdef\gplbacktext{}%
  \gdef\gplfronttext{}%
  \makeatother
  \ifGPblacktext
    \def\colorrgb#1{}%
    \def\colorgray#1{}%
  \else
    \ifGPcolor
      \def\colorrgb#1{\color[rgb]{#1}}%
      \def\colorgray#1{\color[gray]{#1}}%
      \expandafter\def\csname LTw\endcsname{\color{white}}%
      \expandafter\def\csname LTb\endcsname{\color{black}}%
      \expandafter\def\csname LTa\endcsname{\color{black}}%
      \expandafter\def\csname LT0\endcsname{\color[rgb]{1,0,0}}%
      \expandafter\def\csname LT1\endcsname{\color[rgb]{0,1,0}}%
      \expandafter\def\csname LT2\endcsname{\color[rgb]{0,0,1}}%
      \expandafter\def\csname LT3\endcsname{\color[rgb]{1,0,1}}%
      \expandafter\def\csname LT4\endcsname{\color[rgb]{0,1,1}}%
      \expandafter\def\csname LT5\endcsname{\color[rgb]{1,1,0}}%
      \expandafter\def\csname LT6\endcsname{\color[rgb]{0,0,0}}%
      \expandafter\def\csname LT7\endcsname{\color[rgb]{1,0.3,0}}%
      \expandafter\def\csname LT8\endcsname{\color[rgb]{0.5,0.5,0.5}}%
    \else
      \def\colorrgb#1{\color{black}}%
      \def\colorgray#1{\color[gray]{#1}}%
      \expandafter\def\csname LTw\endcsname{\color{white}}%
      \expandafter\def\csname LTb\endcsname{\color{black}}%
      \expandafter\def\csname LTa\endcsname{\color{black}}%
      \expandafter\def\csname LT0\endcsname{\color{black}}%
      \expandafter\def\csname LT1\endcsname{\color{black}}%
      \expandafter\def\csname LT2\endcsname{\color{black}}%
      \expandafter\def\csname LT3\endcsname{\color{black}}%
      \expandafter\def\csname LT4\endcsname{\color{black}}%
      \expandafter\def\csname LT5\endcsname{\color{black}}%
      \expandafter\def\csname LT6\endcsname{\color{black}}%
      \expandafter\def\csname LT7\endcsname{\color{black}}%
      \expandafter\def\csname LT8\endcsname{\color{black}}%
    \fi
  \fi
  \setlength{\unitlength}{0.0500bp}%
  \begin{picture}(4534.00,3174.00)%
    \gplgaddtomacro\gplbacktext{%
      \csname LTb\endcsname%
      \put(858,594){\makebox(0,0)[r]{\strut{} 0.01}}%
      \put(858,1057){\makebox(0,0)[r]{\strut{} 0.1}}%
      \put(858,1520){\makebox(0,0)[r]{\strut{} 1}}%
      \put(858,1983){\makebox(0,0)[r]{\strut{} 10}}%
      \put(858,2446){\makebox(0,0)[r]{\strut{} 100}}%
      \put(858,2909){\makebox(0,0)[r]{\strut{} 1000}}%
      \put(990,374){\makebox(0,0){\strut{} 4}}%
      \put(1515,374){\makebox(0,0){\strut{} 5}}%
      \put(2039,374){\makebox(0,0){\strut{} 6}}%
      \put(2564,374){\makebox(0,0){\strut{} 7}}%
      \put(3088,374){\makebox(0,0){\strut{} 8}}%
      \put(3613,374){\makebox(0,0){\strut{} 9}}%
      \put(4137,374){\makebox(0,0){\strut{} 10}}%
      \put(220,1751){\rotatebox{-270}{\makebox(0,0){\strut{}$n_0$ ($10^{12}\,\mathrm{m}^{-3}$)}}}%
      \put(2563,154){\makebox(0,0){\strut{}E (MV/m)}}%
    }%
    \gplgaddtomacro\gplfronttext{%
      \csname LTb\endcsname%
      \put(1782,2736){\makebox(0,0)[r]{\strut{}$k = 1$}}%
      \csname LTb\endcsname%
      \put(1782,2516){\makebox(0,0)[r]{\strut{}$k = 2$}}%
      \csname LTb\endcsname%
      \put(1782,2296){\makebox(0,0)[r]{\strut{}$k = 3$}}%
    }%
    \gplbacktext
    \put(0,0){\includegraphics{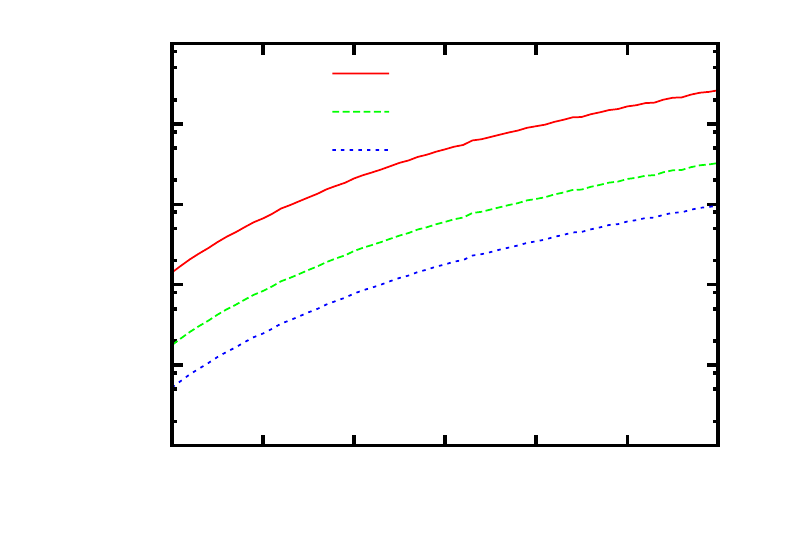}}%
    \gplfronttext
  \end{picture}%
\endgroup